\documentclass[aps,preprint]{revtex4}

\usepackage{graphicx}
\def\be{\begin{equation}}
\def\ee{\end{equation}}
\def\ba{\begin{eqnarray}}
\def\ea{\end{eqnarray}}
\def\bq{\begin{quote}}
\def\eq{\end{quote}}

\begin{document}

\bibliographystyle{apsrev}

\preprint{SU-GP-01/11-1}

\title{Early Universe Cosmology and Tests of Fundamental Physics:\\
Report of the P4.8 Working Subgroup, Snowmass 2001}



\author{Andreas Albrecht$^1$, Joshua A. Frieman$^{2,3}$, Mark Trodden$^4$
\\ }
\affiliation{$^1$Department of Physics \\
University of California at Davis \\
One Shields Ave. \\
Davis, CA 95616
\\ \\
$^2$NASA/Fermilab Astrophysics Center\\
Fermi National Accelerator Laboratory\\
P. O. Box 500, Batavia, IL 60510 \\ \\
$^3$Department of Astronomy \& Astrophysics \\
The University of Chicago \\
5640 S. Ellis Avenue, Chicago, IL 60637\\ \\
$^4$Department of Physics \\
Syracuse University \\
Syracuse, NY, 13244-1130\\}


\date{\today}

\begin{abstract}
This is the report of the {\em Working Group on Early Universe Cosmology
and tests of Fundamental Physics}, group P4.8 of the of the Snowmass 2001
conference.  Here we summarize the impressive array of advances that
have taken place in this field, and identify opportunities for even
greater progress in the future.  Topics include Dark Energy, Cosmic
Acceleration, Inflation, Phase Transitions, Baryogenesis, and
String/M-theory Cosmology.  The introductory section gives an
executive summary with six key open questions on which we can expect
to make significant progress.
\end{abstract}

\maketitle

\section{Introduction}
\label{P48Intro}

Although it is perhaps presumptious to judge the long-term significance of
events without the remove of history, it is widely accepted that
cosmology is now undergoing a renaissance.
An impressive body of new data from a variety of observatories and experiments
is arriving every year, and exciting theoretical
ideas are being put to the test.
A major driving force behind these
dramatic developments has been the application of ideas from
particle physics to the early Universe. These ideas have
resulted in
concrete proposals for the state and matter content of the Universe
today, allowing the new data to take root in a rich context of
fundamental physics.  Thanks to this theoretical framework, the vast new
datasets on the cosmic microwave background anisotropy and large-scale
structure provide much more than brilliant cartography:  they
are addressing deep questions about the nature of matter,
space, and time.  In fact, the exciting opportunities presented
by links to the early Universe and fundamental physics provided
essential motivation for collecting the new data in the first place.

The excitement in this field was clearly evident in the sessions
devoted to the early Universe and fundamental physics at Snowmass
2001.  The high energy physics community can take pride in its key
role in stimulating such advances in cosmology and
celebrate the insights into fundamental physics that have already
emerged from this activity.
Here we outline the status of and recent advances in early Universe
cosmology (interpreting this phrase quite broadly) and document the
abundant opportunities for future progress.  The combination of an
impressive track record and great future opportunities makes a strong
case for continuing the high energy physics community's role as a
driving force in the field of cosmology.
We can do this by exploiting the existing opportunities at the
interface between particle physics and cosmology and by vigorously
pursuing the fundamental questions that are central to particle
physics.  Advances on these frontiers are bound to create more
opportunities to shape the future of cosmology and reap even greater
rewards in the form of further insights into fundamental physics.

Over the next decade, the cosmological parameters determining the structure of
the Universe will be determined to within a few percent. This era
of precision cosmology will bring sharply into focus the issues of
fundamental physics underlying the values of these parameters. The
focus of our working subgroup was to survey and evaluate the
status of early Universe cosmology, to consider how upcoming data,
both astrophysical and collider-based, will shape our knowledge of
the earliest times in the Universe, and to develop conclusions
about the most promising avenues of research.

We have organized this report into six sections: dark matter; dark energy and
the accelerating
universe; inflation; cosmic phase transitions;
baryogenesis; and cosmology and fundamental physics. Most
of these topics are deeply intertwined with one another, so many
specific issues turn up in more than one section.
Before turning to these topics in depth, we briefly summarize their
status with a series of open questions:

\begin{itemize}

\item {\bf What is the Dark Matter?} While there is now compelling evidence
that 30\% of the critical density of the Universe is in the form of
non-baryonic dark matter, and particle physics beyond the standard model
provides attractive candidates, we have no direct evidence about its
identity. Direct and indirect dark matter searches, including
accelerator-based experiments, will be critical for
helping unravel the identity of the dark matter.

\item {\bf What is the nature of the Dark Energy?} While there are now
multiple lines of evidence indicating that 70\% of the critical density of
the Universe is in the form of a negative-pressure Dark Energy component,
we have no firm clues as to its origin and nature. Theoretical studies
operate in the shadow of the cosmological constant problem, the most
embarrassing hierarchy problem in particle physics. Experiments to be
carried out over the next decade should shed considerable light on the
matter, by constraining the Dark Energy equation of state and determining
whether it is consistent with vacuum energy or something else.

\item {\bf Did inflation occur in the early Universe?} Inflation provides
the only well-studied paradigm for explaining the observed homogeneity and
inhomogeneity in the Universe, but a consensus model has not been developed.
In the near term, CMB anisotropy experiments will probe inflation via
precision tests of cosmological parameters. In the longer term,
polarization experiments hold out the prospect of perhaps observing
the imprint of inflation-generated gravitational waves.

\item {\bf Are there observable relics from cosmic phase transitions?}
Phase transitions may play a role in
generating a variety of phenomena, from topological
defects to baryogenesis to dark matter, magnetic fields, and ultra-high
energy cosmic rays.

\item {\bf Why is there more matter than antimatter in the Universe?}
While the necessary
ingredients for successful baryognesis have been known for over 30 years,
there is no consensus model for the actual mechanism. Recent attention
has focused on baryogenesis at the electroweak scale, an idea which
may be tested indirectly by accelerator experiments.

\item{\bf What roles do string theory, quantum gravity, and extra
dimensions play in cosmology?} Perhaps subsantial ones, but
current investigations have been mainly limited to toy models in
the absence of a well-understood fundamental theory. Recent attention
has focused on scenarios in which standard model particles are
confined to a brane in a Universe with large extra dimensions.

\end{itemize}



\section{Dark Matter}
\label{P48DM}
Over the last thirty years, a mountain of evidence has accumulated
indicating that the bulk of the matter in the Universe is {\it dark}.
Observations of galaxies and galaxy clusters
reveal substantially more mass associated
with these systems than can be attributed to
the luminous matter. The evidence includes
flat spiral galaxy rotation curves, dynamical studies of satellite galaxies,
galaxy-galaxy lensing,
dynamical, X-ray, weak lensing, and Sunyaev Zel'dovich (SZ)
studies of galaxy
clusters, and the large-scale peculiar motions of galaxies. Taken together,
these observations have consistently pointed to a
matter density, expressed as a fraction of the critical density for a flat
Universe, of $\Omega_m \simeq 0.15 - 0.3$. In addition,
recent measurements of
the cosmic microwave background (CMB) anisotropy, of the cosmic shear
(large-scale weak lensing), of the galaxy and Lyman-alpha forest
clustering power spectra, and of the galaxy cluster abundance
have provided independent and consistent
estimates of the cosmic mass density (in the
context of additional assumptions about the formation of structure in
the Universe and in combination with measurements of the Hubble parameter),
yielding $\Omega_m \simeq 0.3$ (for recent reviews, see, e.g.,
\cite{Primack:2000dd, Turner:2001mw}).

The bulk of the dark matter in the Universe
must be non-baryonic. Estimates of the cosmic
baryon density, traditionally from big bang nucleosynthesis
(e.g., \cite{Burles:2000zk}) and
more recently from the CMB anisotropy
\cite{Pryke:2001yz, Netterfield:2001yq, Stompor2001},
now yield $\Omega_b h^2 = 0.02$;
combined with measurement of the Hubble parameter ($h=0.72 \pm 0.07$ from
the HST Key Project \cite{Freedman:2001}),
this implies $\Omega_b \simeq 0.04$, substantially
below the total matter density. The case
for non-baryonic dark matter
has been strengthened by independent measurements
of $\Omega_m/\Omega_b \simeq 7-9$, from the cluster baryon fraction
(SZ and X-ray measurements) and from preliminary detection of baryonic
wiggles in the large-scale power spectrum \cite{Percival2001, Miller:2001cf}.
In addition, the fact that $\Omega_b > \Omega_{Luminous} \simeq 0.007$
argues that a substantial fraction of the baryons in the Universe
are also dark (perhaps in the form of compact objects or MACHOs).

Since the pattern of CMB anisotropy indicates
that the spatial geometry of the universe is nearly flat, $\Omega_{tot}
\simeq 1$ \cite{Pryke:2001yz, Netterfield:2001yq, Stompor2001},
the Universe must be dominated by a component---the so-called Dark
Energy---which is smoothly
distributed on at least the scale of clusters. In order for this component
not to have disrupted the formation of structure, it should have
come to dominate the energy density only at quite recent epochs, which
implies that its effective pressure should be negative.
This is consistent with observations of the apparent brightness
of high-redshift SNe Ia, which indicate directly the presence of
dark energy accelerating the Universe \cite{Perlmutter, Riess}.
A consistent cosmological model has thus emerged, in which
$\Omega_m \simeq 0.3$ and $\Omega_{de} \simeq 0.7$.

While the observational evidence for dark matter and dark energy
has been building, we still have no solid clues as
to the identities of either of
these two components. Nevertheless, their mere existence strongly points
to physics beyond (perhaps way beyond) the standard model. Experiments
aimed at trying to discover the nature of the dark matter and the
dark energy are therefore critical for progress in
both particle physics and cosmology. For the remainder of this
section, we focus on recent developments in understanding the role
of dark matter; the following section describes dark energy.

As has often been
pointed out, the uncertainty in the mass of the (non-baryonic)
dark matter constituent ranges over at least
70 orders of magnitude, from $\sim 10^{-5}$ eV (for axions)
to $\sim 10^{63}$ eV (for planetary mass primordial black holes).
Within this vast range, the theory of structure formation
provides indirect evidence about some of the properties which
the (bulk of the) dark matter must have (see below).

From the theoretical perspective, particle physics theories beyond
the standard model do provide well-motivated candidates for
non-baryonic dark matter. In supersymmetric models with conserved
R-parity, the lightest supersymmetric partner (LSP) of ordinary
fermions and bosons is stable. Such a particle is weakly
interacting and has a mass of order the electroweak scale (hence
the moniker WIMP, for weakly interacting massive particle); in
combination, these properties imply that the LSP should have a
relic cosmic density of order $\Omega_m \sim 1$ (within a few
orders of magnitude) \cite{Jungman:1996df}. The axion, a stable
pseudo-Nambu-Goldstone boson which emerges from models which
address the strong CP problem via a global $U(1)$ (Peccei-Quinn)
symmetry, is also constrained by astrophysical and cosmological
arguments to have a density of order the critical density, if it
exists.

The SUSY LSP and the axion are both plausible
candidates for cold dark
matter, with similar effects on the growth of large-scale structure,
but their experimental signatures are quite different
\cite{Sadoulet:1999rq}. Direct searches
for WIMPs in the halo of the Galaxy are now becoming mature---relying
on the deposition of $\sim$ keV of recoil energy when a WIMP scatters
from a nucleus in a detector. Several experiments have reported bounds
on WIMP masses and cross-sections \cite{Abusaidi:2000wg,
Baudis:2000ph},
with one controversial report of
a detection via the annual modulation signal \cite{Bernabei:2000qi}.
The challenge for the next generation of
direct detection experiments is to scale up the detector mass while
continuing to beat down systematic backgrounds, in order
to achieve sensitivity to much smaller
event rates and thereby probe a large swath of SUSY parameter space.
In addition, indirect WIMP searches, which rely on detection of
high-energy gamma rays or charged particles from WIMP annihilation in the
halo, or high-energy neutrinos from annihilations in the Earth or the Sun,
will gain sensitivity with the coming round of large experiments such
as GLAST, VERITAS, and ICECUBE. These direct and indirect
WIMP searches should be considered
complementary to searches for supersymmetry at colliders.
Axion searches involve the resonant conversion of halo axions into
microwave photons in the presence of a strong magnetic field; several
experiments around the world are underway and are also planning upgrades
which should enable them to probe the range of axion masses and
couplings expected from theory \cite{Asztalos:2001tf, Asztalos:2001jk}.
Both direct and indirect searches for particle dark matter are
sensitive to some degree to the phase space distribution of dark matter
particles in the Galaxy halo. Recent N-body simulations of cold dark
matter have stimulated investigations of the expected
clumpy nature of halo dark matter and its
possible implications for experimental signatures \cite{Blasi:2000ud,
Stiff:2001dq, Moore:2001vq, Calcaneo-Roldan:2000yt}.

In assessing dark matter candidates, we should continue to be
cognizant of possible surprises and therefore keep an open mind: theory
has provided a multitude of
possible candidates beyond WIMPs and axions and could provide new
ones. The experiments above are rightly aimed at what are currently
considered the most plausible theoretical candidates, but some
thought should go into constraining other possibilities.

Models of structure
formation provide important clues about the nature of the dark matter,
strongly suggesting that the bulk of it is (at most) weakly
interacting and non-relativistic at
late times (cold dark matter, CDM). Measurements of the CMB anisotropy
on degree scales and larger indicate that the inflationary paradigm
with nearly scale-invariant, adiabatic perturbations is a very strong
candidate for the origin of structure. As noted above,
in the context of this paradigm, measurements of galaxy and mass
clustering from galaxy surveys point to a model
with $\Omega_m \simeq 0.3$ in a dark matter component which
can freely cluster on scales larger than of order a few
kpc. On the other hand, as cosmological
N-body simulations of structure formation
have pushed to resolve smaller scales, they have uncovered potential
discrepancies between CDM models, in which the
dark matter is assumed cold (non-relativistic) and collisionless
(weakly interacting with itself and with baryons and photons), and the
observed properties of galaxy halos. In particular,
CDM models predict dark matter halos with steep, `cuspy'
inner density profiles, $\rho(r) \sim r^{-n}$,
with $n \simeq 1 - 1.5$, while rotation curves for dwarf and
low surface brightness (LSB) galaxies indicate constant density
cores (e.g., \cite{Flores:1994gz, Moore:1994wb, Navarro:1996iw,
Moore:1999gc, Jing99, Klypin2000}).
In addition, these simulations predict that the Local Group of
galaxies should include substantially more
dwarf satellite galaxies than are
observed \cite{kauffmann93, Moore:1999wf, Klypin:1999uc}: CDM halos
appear to have too much surviving substructure.

The cusp and substructure problems (among others
\cite{Sellwood:2000wk})
have prompted a number of authors to
recently (re-)consider scenarios
in which the fundamental properties of the dark matter
are modified. In these alternatives, one no longer assumes that
(all) the dark matter is both cold and collisionless:
it has a new property which suppresses its
small-scale clustering, thereby causing halos to be less cuspy and lumpy.
Examples include dark matter which
self-interacts \cite{Spergel:1999mh, 2000NewA....5..103G,
Peebles:1999se},
annihilates \cite{Riotto:2000kh, Kaplinghat:2000vt},
decays \cite{2001ApJ...546L..77C},
has a non-negligible Compton wavelength (fuzzy dark matter)
\cite{Hu:2000ti},
or has a non-negligible velocity dispersion (warm dark matter)
\cite{1996ApJ...458....1C, 2001ApJ...551..608S, 2001ApJ...556...93B}.
Another possibility is to stick with cold, collisionless dark matter
but suppress the primordial power spectrum on small length scales
\cite{Kamionkowski:1999vp, 2000ApJ...539..497W}.

The degree to which these different alternatives solve the
difficulties of `oridinary' cold dark matter has been somewhat
controversial \cite{2000ApJ...544L..87Y, miralda2000,
2001ApJ...547..574D, dalc2000, 2000ApJ...543..514K}. From the
theoretical standpoint, the proposed new dark matter properties
are not particularly attractive. For example, warm dark matter
requires a stable particle with a mass of order 1 keV, not
particularly close to the electroweak or SUSY scale, which must
decouple before a significant amount of entropy is transferred to
the CMB, so that its cosmic abundance can be suppressed. For
annihilating dark matter, one must suppress catastrophic
annihilations in the early universe. For self-interacting dark
matter, one must supply a new interaction with the requisite
strength.  The case for `non-standard' dark matter properties
would certainly be more appealing if they could be shown to arise
naturally in the context of compelling extensions of the standard
model of particle physics; while some work has been done along
these lines (e.g., \cite{Bento:2000ah, 2001ApJ...551..608S}), this
remains a challenge for model-builders.

It should also be noted that there may be
more pedestrian `astrophysical' explanations for these discrepancies,
involving either the data or the fact that the simulations include only a
limited physical description of the baryons. For example, new and reanalyzed
data on the rotation curves of dwarf and LSB galaxies, with
allowance made for beam-smearing effects,  has led some authors to
conclude that these systems do not discriminate
strongly between constant density and cuspy inner
halos \cite{2000AJ....119.1579V,
2001MNRAS.325.1017V, 2001gddg.conf..545S}; however, another recent study
has found that LSB rotation curves are definitely {\it not} well fit by cuspy
cores \cite{2001ApJ...552L..23D}.
It has also been suggested that the interactions of
supermassive black holes
(now known to be ubiquitous in the cores of galaxies)
could destroy dark matter cusps when young galaxies merge
\cite{2001ApJ...551L..41M}.
In addition, the overabundance of galactic satellites may be
reduced by reionization,
which  suppresses gas accretion and thus
star formation in these low-mass clumps \cite{2000ApJ...539..517B}. In
this picture, the observed lack of halo substructure may be a property of the
stellar baryons but not of the dark matter. Finally, it has been
suggested that both the cusp and substructure problems could
be resolved by the effects of galactic winds \cite{2001MNRAS.321..471B}.

More data on the structure of halos and improved modeling of them is needed
to ultimately resolve whether observed galaxy halos are
consistent with `ordinary' cold dark matter. Nevertheless,
the study of alternative dark matter properties that the
cusp and substructure problems stimulated remains of interest,
because the issue can be turned around: we can use
structure formation to constrain the properties of dark matter
\cite{Hogan:2000bv}. For example, for warm
dark matter (WDM), the high phase space density of dwarf spheroidal
galaxies implies a lower limit on the WDM particle mass, $m_X > 0.7$ keV
\cite{dalc2000}. The observed opacity distribution of the Lyman-alpha
forest at redshift $z \sim 3$ leads to a similar lower mass limit
\cite{2000ApJ...543L.103N}. Requiring that sufficiently massive
black holes be able to form in time to power the observed highest
redshift quasars at $z \sim 6$ and that high-redshift galaxies be
able to reionize the Universe by that epoch also lead to qualitatively
similar bounds \cite{bark2001}. On the other hand, if the WDM mass
is much above 1 keV, it will only suppress power on mass scales
well below $10^{10} M_\odot$ and therefore lead to structure on
galaxy scales that is indistinguishable from CDM.
Further data on halo structure, e.g., from strong gravitational lensing \cite{
keet2001, 2001ApJ...549L..25K} and from galaxy-galaxy lensing,
and on halo clustering and abundances at high redshift
will help constrain the nature of the dark matter.
(For example, self-interacting dark matter models generally predict
that galaxy halos are spherical instead of elliptical; in principle, the
shapes of halos can be probed by lensing, by the dynamics of halo
tracers in the Galaxy, by polar ring galaxies, and by X-rays from
massive galaxies, among other methods.)

Large-scale structure can also place
useful constraints on the masses of
particles which contribute only a small
fraction of the dark matter density---neutrinos. The atmospheric
neutrino data from Super-Kamiokande and MACRO, interpreted as an effect of
neutrino oscillations, indicate a neutrino mass squared difference
of order $\delta m^2 \simeq (2-6)\times 10^{-3}$ eV$^2$,
which implies a lower bound on the neutrino cosmic density of
$\Omega_\nu > 0.0008$. On the other hand, the
observed clustering of galaxies and the Lyman-$\alpha$ forest
implies an upper bound on $\Omega_\nu$: since neutrinos are
relativistic until late times, they free-stream out of
perturbations on small scales, thereby damping small-scale power if
they make an appreciable contribution to $\Omega_m$. The current
observations translate into
the (roughly $2\sigma$) upper limit $m_\nu < 3$ eV for the combined masses of
light stable neutrinos \cite{Croft:1999mm, wang2001},
comparable to current experimental
limits on $m_{\nu_e}$ from tritium experiments. In the near future,
neutrino masses as low as $m_\nu \sim 0.3$ eV can be probed by combining
CMB experiments (MAP and Planck) with galaxy and Lyman-$\alpha$ forest
power spectrum data from the Sloan
Digital Sky Survey \cite{Hu:1998mj}.
These improved constraints are again comparable to
expected improvements in the experimental bounds on $m_{\nu_e}$.

Partly motivated by the perceived problems of `ordinary' cold dark
matter noted above, there has been renewed attention paid to
alternatives to dark
matter: the mass discrepancies in galaxies normally ascribed to dark matter
could instead be signalling the breakdown of Newtonian gravity (for a
recent review, see, e.g., \cite{Sellwood:2000wk}).
Until the dark matter is actually detected, this may
remain a logical possibility.
The most commonly discussed alternative,
modified Newtonian dynamics (MOND) \cite{1983ApJ...270..365M},
may be expressed as a
modification of the law of inertia below some fundamental
acceleration scale; with an appropriate modification,
the observed flat rotation curves of galaxies can be reproduced
\cite{1983ApJ...270..371M}.
The degree to which MOND is consistent with the range of
astrophysical data continues to be debated. Moreover, the fact that
MOND is only a phenomenological prescription
for describing dynamical systems, not a fundamental theory, has
hampered attempts to apply it to cosmology \cite{Scott:2001td},
structure formation, and gravitational lensing \cite{mort01}.

While it is important to
keep an open mind to dark matter alternatives, it is
also necessary to subject them to observational tests and to hold
them up to the lamp of theoretical plausibility. When MOND was first
proposed, in the early 1980's, galaxy rotation curves offered the
primary evidence for a mass discrepancy, and MOND was aimed at
providing an alternative explanation for these observations. Although
rotation curves still provide the strongest evidence
for a mass discrepancy, ancillary circumstantial evidence for dark matter
has built up substantially
in the intervening years. As noted at the beginning of this Section,
this newer evidence is of two kinds: (a) direct inference of mass
discrepancies in galaxies and clusters using a variety of probes, and
(b) consistency of the cold dark matter model with $\Omega_m \simeq 0.3$
with CMB, SNe Ia, large-scale structure, and weak lensing data.
Although MOND cannot address most of these other observations without being
embedded in a fundamental theory, as these new pieces of evidence
mount up the possibility of explaining them all with something other
than dark matter becomes less likely.
On the theoretical side, while particle physics theory provides
well-motivated candidates for cold dark matter,
it has proved difficult to embed MOND in a more fundamental theory;
part of this difficulty likely traces to the fact that it appears
to violate cherished principles such as the equivalence principle,
Lorentz invariance, and conservation of momentum \cite{Scott:2001td}.
Again, if it
could be shown that dark matter alternatives arise naturally
from new ideas in particle physics or gravitation, the case would
be substantially more compelling (for some attempts, see, e.g.,
\cite{Kinney2000hu,1989ApJ...342..635M}).
Finally, it should be noted that
modified gravity could in principle be falsified (along with
self-interacting dark matter) by better data on the shapes of `dark'
halos or by confirmation of the existence
of dark clumps (several of which have been inferred from weak lensing
observations) \cite{Sellwood:2000b}.

\section{Dark Energy and The Accelerating Universe}
\label{P48DE}

Recent observations of type Ia supernovae (SNe Ia) at high redshift
indicate that the
expansion of the Universe is accelerating \cite{Perlmutter, Riess}:
although concerns about systematic errors remain,
these calibrated `standard' candles appear fainter than would be
expected if the expansion were slowing due to gravity.
According to General Relativity,
accelerated expansion requires a dominant component with effective negative
pressure, $w = p/\rho < 0$.
Such a negative-pressure component is now generically termed
Dark Energy; a cosmological
constant $\Lambda$,
with $p_\Lambda = - \rho_\Lambda$, is the simplest but not the only
possibility.
As noted in the previous Section, recent results for the CMB anisotropy,
which favor a nearly flat Universe, $\Omega_{total} = 1$, coupled with
a variety of observations pointing unambiguously to low values
for the matter density parameter, $\Omega_{m}=0.3$,
provide independent evidence for a dark energy component
with $\Omega _{de }\simeq 0.7$. Such a cosmological model, with
the additional assumptions that the dark matter is (mostly) cold and
that the initial perturbations are adiabatic and nearly scale-invariant
(as predicted by inflation), is in excellent agreement with CMB and
large-scale structure data as well.

On the other hand, the history of
dark energy---more specifically the cosmological constant---is not pretty:
beginning with
Einstein, it has been periodically invoked by cosmologists out of
desperation rather than desire, to reconcile theory with observations,
and then quickly discarded when improved data or interpretation showed
it was not needed. Examples include the first `age crisis'
arising from Hubble's large value for the expansion rate (1929),
the apparent clustering of QSOs at a particular redshift (1967),
early cosmological tests which indicated a negative deceleration
parameter (1974), and the second `age crisis' of the mid-1990's arising from
new evidence in favor of a high value for the Hubble parameter.
Despite these false starts, it seems more likely that dark energy is finally
here to stay, since we now have multiple lines of evidence pointing to it.

With or without dark energy, a consistent description of the vacuum presents
particle physics with a major challenge: the cosmological constant problem
(see, e.g., \cite{weinberg, carrollpressturner}).
The effective energy density of
the vacuum---the cosmological constant---certainly satisfies
$\Omega_\Lambda < 1$, which
corresponds to a vacuum energy density
$\rho_{\Lambda} =\Lambda/8\pi G < (0.003 \; {\rm eV})^4$.
Within the context of quantum field theory, there is as yet no
understanding of why the vacuum energy density arising from
zero--point fluctuations is not of order the
Planck scale, $M^4_{Pl}$, 120 orders of magnitude larger,
or at least of order the supersymmetry breaking scale, $M^4_{SUSY} \sim {\rm
TeV}^4$, about 50 orders of magnitude larger.
Within the context of classical
field theory, there is no understanding of why the vacuum energy
density is not of the order of the scale of one of the vacuum
condensates, such
as $M^4_{GUT}$, $M^4_{SUSY}$,
$M_W^4\sin^4\theta_W/(4\pi\alpha)^2
\sim (175 \; {\rm GeV})^4$, or $f_\pi^4 \sim (100\; {\rm MeV})^4$.
The observational upper bound on the vacuum density appears to require
cancellation between two (or more) large numbers to very high precision.
Note that this is not an argument against the cosmological constant
{\it per se}, merely a statement of the fact that we do not understand
why $\Lambda$
is as small as it is. Some theorists expect that whatever explains
the smallness of the cosmological constant---e.g., some as yet
undiscovered (or presently misunderstood) symmetry---may require it to
be exactly zero,
but as of yet no compelling mechanism has been proposed.
This discrepancy is the most embarrassing hierarchy problem for
modern particle physics theory. Moreover,
the arguments above indicate that it is manifest even at low energies.
It is therefore not obvious that its solution necessarily lies in unraveling
physics at ultra-high energies, e.g., in string theory.

The cosmological constant problem predates the recent
evidence for dark energy. However,
dark energy raises a new puzzle, the so-called
coincidence problem. If the dark energy
satisfies $\Omega_{de} \simeq 0.7$,
it implies that we are observing the Universe just at the special epoch
when $\Omega_m$ is comparable to $\Omega_{de}$, which might seem to beg for
further explanation. We might rephrase these two problems as follows:
(a) why is the vacuum energy density so much smaller than the
fundamental scale(s) of
physics? and (b) why does the dark energy density
have the particular non-zero value that it does today?
If the dark energy is in fact vacuum energy (i.e., a non-zero
cosmological constant), then the answers to these two questions are
very likely coupled; if the dark energy is not due to a pure
cosmological constant, then these questions may be logically disconnected.

In recent years, a number of models in which the dark energy is
dynamical, e.g., associated with a scalar field and not a fundamental
cosmological constant, have been discussed (e.g.,
\cite{dolgov, ratpee, wetterich, fhsw, Caldwell:1998ii, Binetruy:1998rz,
Zlatev:1998tr,buch, choi,
Masiero:1999sq, Sahni:1999qe, Dodelson:2000fq, Albrecht:1999rm,
Albrecht:2001xt}).
These models, sometimes known as ``quintessence'' models,
start from the assumption that questions (a) and (b)
above are logically disconnected. That is, they postulate that
the fundamental vacuum energy
of the universe is (very nearly) zero, owing to some as yet not understood
mechanism, and that this new physical mechanism
`commutes' with other dynamical effects that lead
to sources  of energy density.  This assumption implies
that all such models do {\it not} address the cosmological constant
problem.
If this simple hypothesis is the case, then the {\it effective}
vacuum energy at any epoch will be dominated by the fields with the
largest potential energy
which have not yet relaxed to their vacuum state. At late times, these
fields must be very light.

Adopting this working hypothesis, we can immediately identify
generic features which a classical model for the dark energy
should have. Dark energy is most simply stored in the
potential energy $V(\phi) = M^4 f(\phi)$ of a scalar field $\phi$,
where $M$ sets the
characteristic height of the potential. The working hypothesis
sets $V(\phi_m)=0$ at the minimum of the potential (although this
assumption is not absolutely necessary); to generate a non-zero
dark energy at the present epoch, $\phi$ must be displaced from the
minimum ($\phi_i \neq \phi_m$ as an initial condition), and to
exhibit negative pressure its kinetic
energy must be small compared to its potential energy.
This implies that the motion of the field
is still relatively damped, $m_\phi = \sqrt{|V''(\phi_i)|}
< 3H_0 = 5\times 10^{-33} h$ eV. Second, for $\Omega_\phi \sim 1$ today,
the potential energy density should be of order the critical density,
$M^4f(\phi) \sim 3H^2_0 M^2_{Pl}/8\pi$, or (for $f \sim 1$)
$M \simeq 3\times 10^{-3}h^{1/2}$ eV; the resulting value of the scalar
field is typically $\phi \sim M_{Pl}$ or even larger.
Thus, the characteristic height and curvature of the potential are
strongly constrained for a classical model of the dark energy.

This argument raises an apparent difficulty for all ``quintessence'' models:
why is the mass scale $m_\phi$ at least
thirty orders of magnitude smaller than $M$
and 60 orders of magnitude smaller than the characteristic field value
$\phi \sim M_{Pl}$?
In quantum field theory, such ultra-low-mass scalars are not {\it generically}
natural: radiative corrections generate large mass renormalizations at
each order of perturbation theory. To incorporate ultra-light scalars
into particle physics, their small masses should be at least
`technically' natural, that is, protected
by symmetries, such that when the small masses are set
to zero, they cannot be generated in any order of perturbation theory,
owing to the restrictive symmetry. While many phenomenological quintessence
models have been proposed, with few exceptions \cite{fhsw, Binetruy:1998rz,
choi, Albrecht:2001xt}, model builders have ignored this
important issue. In many quintessence models, particularly those with
``runaway'' potentials (e.g., exponentials, inverse power laws),
this problem is compounded by the fact that the scalar field amplitude
at late times satisfies $\phi \gg M_{Pl}$: the dark energy dynamics
would be ruined by generically expected terms of
the form $\phi^{4+n}/M^n_{Pl}$ unless they are highly suppressed.
In addition, such `eternal' quintessence models lead to horizons,
which may create difficulties for (perturbative) string theory \cite{dine}.
On the plus side, models in which the ``quintessence'' field is
an axion with the requisite mass scales \cite{fhsw} may
perhaps arise in perturbative string theory \cite{choi, dine}, and
the radion field in brane models with large extra dimensions
may also have the requisite properties for dark energy \cite{Albrecht:2001xt}.

Since the quintessence field must be so light, its Compton wavelength
is comparable to the present Hubble radius or larger. As a result,
depending on its couplings,
it may mediate a new long-range force, a possibility constrained by
equivalence principle experiments and astrophysical tests \cite{carroll}.
While quintessence models have been
constructed which evade these bounds \cite{Albrecht:2001xt}, these constraints
remain an important consideration for theorists in bulding models.
More sensitive equivalence principle experiments are warranted,
as they will provide stronger constraints upon and (possibly) evidence
for dark energy.

While ``quintessence'' models provide theoretical
scenarios of varying plausibility for a dark energy component
which differs from a cosmological constant ($w = p_\phi/\rho_\phi > -1$
if the field is rolling), the next major
developments in this field will likely come from observations:
progress in probing dark energy will be critical
to pointing the way to theoretical understanding of it.
The first challenge for the coming decade is to determine
whether the dark energy equation of state parameter $w$ is
consistent with $-1$ (the vacuum) or not. The current constraints
from SNe Ia observations (e.g., \cite{garnavich,turnwhite2}) are consistent
with $w=-1$, but with large uncertainties. As the errors are
reduced, will $w=-1$ be excluded or preferred? If the latter,
i.e., dark energy consistent with vacuum energy, it will likely be
difficult to make further theoretical progress without tackling
the cosmological constant problem. If the former, i.e., dark energy
inconsistent with vacuum energy, it would appear to point to a dynamical
origin for this phenomenon. In that case, marshalling a variety of
probes, including supernovae, weak lensing, cluster counting
via SZ, Lyman-alpha forest clustering, and others
(discussed elsewhere in these proceedings) to determine $w$
and its possible evolution with redshift will be needed.
Fortunately, it appears
that the prospects for improved SNe observations and the maturation
of complementary
probes are quite good, and with them the prospects for
determining the nature of the dark energy. As the discussion above
indicates, these observations may in some sense be probing physics
near the Planck scale.

\section{Inflation: Models Tests and Alternatives}
\label{P48Inf}

\subsection{Overview}
The idea of cosmic inflation has played a central role in the
development of cosmology over the last 20 years.  Inflation
offers an explanation for many features of the Universe that used to
seem beyond the reach of scientific explanation, in particular
its spatial flatness and homogeneity, and
makes specific predictions for the seeds
that formed galaxies and other structure. (For reviews see for example
\cite{AAreview,LiddleLythBook,LindeBook})

The emergence of the Cosmic Inflation idea has created many exciting
opportunities.   On the astronomical side, it has given us a
``standard model'' with specific values for the density of the Universe,
$\Omega_{tot}$, and the spectrum of deviations from perfect homogeneity,
predictions that can now be confronted with astronomical data.
The prospect of testing
these predictions has been crucial to making the case for the host of
new observational campaigns (such as the Sloan Digital Sky
Survey and the MAP and PLANCK
satellites) which are set to produce a flood of new data.

There are also key unresolved questions within the
inflationary picture.  Progress  on these questions falls squarely in the
domain of high energy physics and is likely to lead to additional
opportunities for observational tests.  One significant question is whether
there is any real competition for cosmic inflation:  are there alternative
dynamical processes that could, like inflation, set up the Universe
with the features we observe?  If so, how can we tell which of
the alternatives
Nature might have actually chosen?  Recent attempts to address
these questions have stimulated considerable interest and have posed
problems for fundamental physics that are exciting in their own right.

\subsection{The basic idea}
Inflation is based on the idea that the Universe could have been
dominated at very early times by an unusual type of matter with an
equation of state $p = w \rho$, with $w < -1/3$.  In most models,
these conditions are achieved by a scalar field $\phi$ (the
``inflaton'') which enters into a state where the potential energy
density ($V(\phi)$) dominates over other terms in the stress-energy
tensor.  Under these conditions, the inflaton has an equation of state
with $w$ close to $-1$, and the Universe enters a period of quasi-exponential
expansion called inflation. In most models, the inflaton evolves
classically down the potential $V$; these are called ``slow roll''
models.

During inflation, the spatial curvature becomes
negligible, leading to a ``flat'' universe (with $\Omega_{tot}$ = 1).
Also, the quasi-exponential
expansion pushes field modes from infinitesimal scales
all the way to the size of the observed Universe and even well
beyond that.  Specific calculations allow us to follow the ``zero
point'' quantum fluctuations in these modes out to cosmic scales and
lead to concrete predictions for the primordial perturbations produced by
a given inflationary scenario.

Cosmic Inflation has predictive power because details of the state of
the Universe before inflation are hidden beyond the domain of
realistic observations.  The observable features of the Universe after
inflation are specified by the dynamics of inflation and are insensitive
to the initial conditions.  To realize this picture, a
minimum number ($N_e$) of $e$-foldings of the scale factor during
inflation  must be achieved
(for example, $N_e \geq 60 $ for inflation at the  Grand
Unification scale).  After a sufficient period of inflation, energy
must be transferred from the (dominant) inflaton field into ordinary
matter via inflaton decay,
causing the Universe to ``reheat''.

A crucial aspect of the
inflationary scenario is that it radically changes the causal
structure of the Universe as compared with the Standard Big Bang.  It
is only thanks to these changes that one can hope to explain the
state of the Universe using causal processes.  Thus, inflation is noted
for ``solving the horizon problem'' (in the sense that it makes a
Universe that appears homogeneous over the present Hubble scale
much more probable, given a variety of initial conditions),
in addition to explaining specific features of the observed Universe.

\subsection{Tests of Inflation}

We first briefly discuss the classic tests of inflation and then continue
with some more subtle issues.  More extended discussion of many
aspects of these tests can be found in the report of the P4.3 group on
{\em CMB and Inflation}.

\subsubsection{Density}
Essentially all models of inflation involve a large amount
of slow-roll inflationary expansion, much more than the minimum $60$
$e$-foldings required. The curvature of the Universe today is
completely negligible in these models, and thus we have the  prediction
$\Omega_{tot} = \bar{\rho}_{tot}/\rho_c \equiv 1$.  However, we only measure
$\bar{\rho}_{tot}$ and $ \rho_c $ in the part of the Universe we can
observe.  Fluctuations in the matter density on the scale of the
present Hubble radius (part of the same spectrum of fluctuations
that produced galaxies and other structure) give the distribution in
the predicted $\Omega_{tot}$ a
small width at the level of one part in 100,000 (much smaller than the
current observational uncertainty).

The current observations on this front are consistent with the
predictions from inflation. Data from the three most recent CMB
experiments give $\Omega_{\rm tot}=1.01 \pm .08, 0.97 \pm .10, 1.0 \pm
.14$ (from DASI~\cite{P48halverson00,P48pryke01},
BOOMERANG~\cite{P48netterfield01}, and MAXIMA~\cite{P48Lee01} respectively).
Combining all these experiments and others~\cite{P48wang01} results
in $\Omega_{\rm tot} = 1.0^{+.06}_{-.05}$.  This result is particularly
impressive since the
largest contribution to $\Omega_{tot}$ comes from the mysterious dark
energy.  Little is known about the nature of the dark matter and
energy of the Universe (see sections \ref{P48DM} and \ref{P48DE} of
this report).  The one sure thing is that it all adds up to
match the inflationary prediction.  Future observations will determine
$\Omega_{tot}$ to higher precision and offer an opportunity to
either confirm or falsify the standard inflationary picture.

\subsubsection{Coherence from inflation}
\begin{figure}[htbp]
\includegraphics[width=3.in]{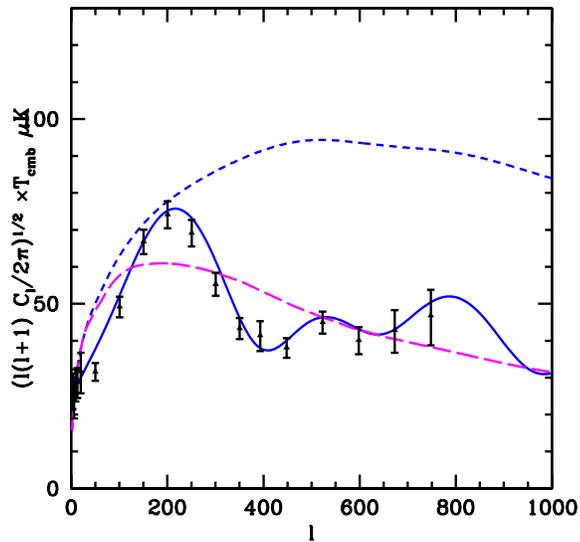}
\caption{Coherence in the CMB: Three models of the Cosmic Microwave
Background anisotropies
are shown with a compilation of the data. The two
dashed curves represent defect models.  The short-dashed curve
is an exotic departure from the standard picture with the
sole motivation of providing a better fit to the data, but even this
curve fails to fit the oscillatory behavior.  The inflation model
(solid curve) with its coherence-related oscillations fits
well. Details of this plot can be found in\cite{Albrecht:2000hu}}
\label{P4_8_EarlyUniverse_fig1}
\end{figure}

As discussed at length by the P4.3 group, inflation gives the
density fluctuations a special property called
``coherence''\cite{1997cdc..conf..265A}.  This
property is related to the dominance of very specific perturbation
modes.  One manifestation of coherence from inflation is a specific
type of oscillation in the spectrum of CMB
anisotropies. Each new round of CMB data has increased the
observational evidence that these oscillations are really there.
Already a class of competing models for the origin of cosmic
structure, the cosmic defect models, have failed largely because they lacked
sufficient coherence to match the data.  Figure
\ref{P4_8_EarlyUniverse_fig1} illustrates this important result.
Future observations will produce much more stringent tests of
coherence and provide an opportunity to support or falsify the
inflationary origin of cosmic structure.

\subsubsection{Gravity waves from inflation}
Perhaps the boldest prediction of the inflationary picture is the
existence of a cosmic gravity wave background (CGB).  There is no known
alternative physical process that would predict anything comparable, so
the detection of this background would be powerful evidence in favor of
inflationary cosmology.  As with the density perturbations,
the gravity wave background is the result of stretching the zero-point quantum
fluctuations in quantum gravitational wave fields (tensor metric
perturbations) to cosmic scales.
Observation of the gravity wave background would provide strong evidence
that the tensor modes of Einstein gravity are quantum mechanical, a
very significant result given the problematic nature of quantum
gravity.  At the
levels predicted by inflation, however, the CGB will be
very hard to detect.  Perhaps the
best hope is through signatures in the polarization of the Cosmic
Microwave Background, as discussed in the P4.3 report.  It seems that a
direct detection of the CGB will remain a challenge for future
generations of cosmologists, but one with very exciting implications.
(Extensive discussion of these and related issues can be found in the
reports of the P4.6 group. A nice summary can be found in
ref. \cite{Hughes:2001ch})

\subsubsection{The scalar spectral index}
As examined at length in the P4.3 report, inflation predicts a
``nearly scale invariant'' spectrum of density perturbations (scalar
metric perturbations).  This corresponds to a ``scalar
spectral index'' $n_s$ for density perturbations of nearly unity.  So
far, the CMB observations are remarkably consistent with this value, and
increasingly tight constraints on the spectral index can be expected
in the near future. The precise nature of the deviations from
scale-invariance depend on the model, but deviations of much more than
20\% are outside the scope of the standard paradigm.

\subsubsection{Further tests of inflation}
As more data comes in, one can begin to take seriously the idea of
making even more detailed tests of the inflationary machinery.
After all, a model of inflation proposes a very specific origin for
the perturbations that formed every observed object in the Universe.
Interesting work has been done on the prospects of actually {\em
reconstructing} the inflaton potential from cosmological data, and the
possibility of other tests is currently an active subject of
investigation. (Further discussion on reconstruction of the inflaton
potential can be found in \cite{Hughes:2001ch,Turner:1996ge})

\subsubsection{Can one {\em really} test inflation?}
A wide variety of models of the Universe have some
kind of inflationary period. A standard paradigm has emerged, which
encompasses the vast majority of existing models.  That is the picture
described so far in this section.  There are a few intriguing
alternative scenarios which incorporate a period of cosmic
inflation but which look very different. For example, there are
models in which our Universe exists inside a single bubble produced in
a cosmic phase transition.  A small amount of inflation is
arranged to happen inside the bubble, and in such models
$\Omega_{tot} \neq 1$ is possible.

So what does it mean to test inflation? The tests described above are
tests of the standard paradigm of inflation.
If all the tests come out positive, the
standard picture will have passed some impressive milestones.  If one
or more tests are negative, the standard picture will have been
falsified, and attention will shift to alternative ideas.  Thus the
field is poised to make dramatic progress.   There is a standard paradigm
which is clearly falsifiable by experiments that are well within reach.

Occasionally there are debates about whether testing the standard
picture described above is truly
testing {\em inflation}, because exotic alternatives
exist. As always, what is tested by observations is determined
by the nature of the observations, not by abstract philosophical
debates.  We urge the community to not let these  debates of principle
detract from the fact that there is very exciting progress to be made.

\subsection{Inflation and fundamental physics}

The mechanism of cosmic inflation explores brand new territory, and
there are many aspects of this idea that need to be better
understood.  In some cases simple assumptions have been made that need
to be justified; in other cases potentially problematic issues have
been ignored, for lack of any concrete way forward.  In addition, a lot of
details are still missing, because we still do not have a consensus
model of  physics on the energy scales at which inflation is supposed to
have happened.  All these issues are linked with fundamental questions
in particle physics.

{\bf Small parameters:}
If a given inflationary model is sufficiently well specified, exact
predictions can be made for the spectrum of cosmic perturbations that
are produced.  In most models, in order to achieve a suitable overall
amplitude for the cosmic perturbations, a dimensionless parameter
in the model must be set to a small value of order $10
^{-12}$.  Although there are claims that in some cases the right
amplitude comes out naturally\cite{Freese:1990rb,Banks:1999ay},
this issue is far
from settled.  We hope that progress on a fundamental description of
physics at high energies will yield a more solid foundation for the inflaton.

{\bf Re/Pre-heating:}
The decay of the inflaton into ordinary matter is a new territory in
its own right.  To have all the energy in the Universe tied up in the
potential energy of a single coherent field and which then decays into
ordinary matter is not yet a well-understood process.  Much of the
analysis has been based on very simple arguments, although some
intriguing coherence effects dubbed ``pre-heating'' have been
investigated (for some recent discussions see
\cite{Boyanovsky:2001va,Kofman:2001rb}.  No doubt there is room for
more progress to be made
in this area, which could be crucial in determining which
inflation models are really viable.  There could be suprises
(i.e., super-efficient or inefficient reheating) which would lead
to very different inflationary scenarios.

{\bf The problem of negative pressure:}
Inflation depends on the inflaton achieving an equation of state with
$p < -\rho/3$. While it is easy enough to construct a scalar field
which has these properties under the right conditions, until
recently it was thought that such states had never been observed in
Nature.  With the discovery of the cosmic acceleration (see
Section \ref{P48DE}) there is evidence that somehow Nature is able to
endow matter with a suitable equation of state, but we are still not
sure how.  In fact, there is even a threat hanging over inflation
related to the cosmological constant problem (discussed in Section
\ref{P48DE}), since the inflaton behaves very much like a cosmological
constant during inflation.  Whatever mechanism Nature chooses to
remove the ``vacuum energy'' (which
naively should exceed observational bounds by 120 orders of
magnitude) could just as well kick in to prevent inflation from
taking place at all.  Alternatively it has been proposed that even ordinary
gravity actually {\em has} a built-in mechanism that can cancel the
vacuum energy with quantum corrections, but that these dynamical
corrections happen slowly enough that it is still possible for a
suitable period of inflation to take
place\cite{Abramo:2001dd,Tsamis:1996qq}.  Whatever the outcome, it is
intriguing that this problem is now linked with the observed
cosmic acceleration, the understanding of which
which there is hope for real progress based on
observations.

{\bf ``Trans-Planckian'' modes and inflation:} During inflation,
quantum field modes are stretched from tiny scales (smaller than the
Planck length) to cosmic scales. What do we really
know about physics on trans-Planck scales? A simple ``Bunch-Davies
vacuum'' provides the required input in the standard calculation,
and it certainly serves the purpose. Interesting recent work
\cite{Martin:2000xs} suggests that only extreme deviations from
the assumed dispersion relation at these scales
could change the predictions for large-scale structure.
Ultimately we would like to see this subject on a much firmer
footing, especially since the prediction of cosmic perturbations
depends on what is input in the first place.  (See the discussion
in subsection \ref{P48TransPlanck}.)

{\bf Before Inflation:}
One of the impressive features of the inflationary picture is
how a period of inflation transforms many different possible initial
conditions into the kind of state we need to kick off the Standard Big
Bang cosmology. It is tempting to think that with that kind of
dynamics, we really do not need to think much about what might have happened
before inflation.  This may in the end turn out to be true, but this
is currently an extremely poorly understood subject, and less pleasing
results may emerge from a more sophisticated treatment.  It is a
challenge to treat quantitatively the ``space of all
pre-inflation states''.  It has even been argued that fundamental
uncertainties to do with placing  measures on pre-inflation states make
predictions from inflation impossible\cite{Linde:1994xx}, but few have
found these arguments compelling (see for example
ref. \cite{Vanchurin:1999iv} for an alternative perspective).  Recent
work has also argued that one cannot have a past described purely by
inflation\cite{Borde:2001nh}, so we are stuck trying to come to
grips with the issue of ``pre-inflation''.  Another approach to this
issue is to make a specific proposal for the ``wavefunction of the
Universe''\cite{Hartle:1983ai,Garriga:1997gr,Vilenkin:1998dn,Hawking:1998bn}
which at least in principle might address these
questions.

\subsection{Alternatives to inflation}
It is quite striking that we have a theory of initial conditions for
the Universe, especially one that is testable and has met with some
success.  But the best way to measure the success of an idea is to
have some real competition.  So far the competition for inflation has
been limited.  (In the more narrow domain of the origin of
perturbations, inflation has already vanquished a class of worthy
competitors, the cosmic defect models\cite{Albrecht:2000hu}.)

But competition does exist: one proposed alternative involves varying
the speed of light (rather than the cosmic expansion) to resolve the
horizon and other
problems\cite{Moffat:1993ud,Albrecht:1998ir},
but this idea still has to find
a compelling foundation in fundamental physics (interesting efforts in
this direction are
ongoing\cite{Alexander:2001ck,Moffat:2001sf}. Another idea is
connected with holography (the
notion that the degrees of freedom of a
gravitating system are much fewer than basic field theory suggests).
It has been proposed that these limitations actually force the early
universe to be highly
homogeneous\cite{Fischler:1998st,Kaloper:1999tt}. But this idea has
yet to take a
concete form with any real predictive power.  There is also the
``ekpyrotic'' scenario\cite{P48ekpyrotic}, which creates the start of
the Big Bang as a
collision between two ``branes'' in a higher dimensional space.  This
picture takes a very different view of explaining initial conditions
from inflation.   Rather than creating a situation in which many
different initial conditions evolve dynamically into what we
need, the ekpyrotic Universe needs to set up extremely special
initial conditions to start with.  But regardless of one's opinion of
this alternative approach, efforts to explore the ekpyrotic scenario
have led to exciting investigations into the nature of colliding
branes\cite{Khoury:2001zk}, work which may
well elucidate interesting issues in brane physics.

\section{Phase transition and their relics}
\label{P48PT}

\subsection{Overview}
\label{P48PTOver}
Spontaneously broken symmetries play a key role in elementary particle
physics.  All the known fundamental forces
of nature (except for gravity) can be described by renormalizable
gauge theories, and the only viable mechanism for giving matter
masses in these theories is spontaneous symmetry breaking.  In
almost all theories with spontaneous symmetry breaking the symmetry is
restored at the high temperatures of the early Universe.  This results
in a cosmological phase transition as the Universe cools through
the critical temperature at which the symmetry becomes broken
\cite{Linde:1979px,VilenkinShellard94,Kibble:1980mv}.

Over much of its history, the evolution of the Universe appears to
have been ``adiabatic'', with matter in local thermal equilibrium.
However, phase
transitions often involve very long timescale processes that can drop
out of local equilibrium and lead to interesting effects.  For example
one can have domain formation, as local regions make different choices
of symmetry breaking direction.
Topological defects such as domain walls, magnetic
monopoles, or cosmic strings form where domains meet. Complete
equilibrium is only achieved when the domains grow (or ``coarsen'') until the
Universe is covered by a single domain, but the coarsening process can
take longer than the present age of the Universe to complete.  Thus
there can be out-of-equilibrium processes that continue right through
to the present day.  Topological defects typically have interior
energy densities
similar to the ambient energy density when they were first
formed, even after the surrounding matter density has dropped by many orders
of magnitude due to cosmic expansion.  Thus defects can
preserve a region with the high densities of the very early Universe
to the present day, offering a unique window on ultra-high
energy physics.

With or without the formation of long-lived defects, the
out-of-equilibrium processes in cosmic phase transitions can lead to a
wide variety of interesting effects. In some cases these effects
introduce exciting new possibilities into the field of cosmology.
In other cases the results of phase transitions are in clear conflict
with observations, firmly ruling out any model that has that type of
transition. The notorious  ``monopole problem'' \cite{Preskill:1979zi} ruled
out almost all models of Grand Unification that were popular at that
time.  Guth's studies of the very same phase transitions led to
his seminal paper on cosmic inflation\cite{Guth:1981zm}.

Cosmic phase transitions could have had a variety of important roles,
from creating baryon number, to producing high energy cosmic rays,
``wimp-zillas'', and a potentially observable background of
gravitational radiation.  For a time, they provided a viable competing
picture for the origin of cosmic structure.  In this section we review
the current status and future opportunities.

\subsection{The many roles of cosmic phase transitions}

{\bf Phase transitions and baryogenesis}
One of the insights we hope to gain from the application of
particle physics to the early Universe is an explanation of the
observed baryon asymmetry of the Universe.  A crucial ingredient in
any baryogenesis scenario is a period during which the relevant processes are
out of equilibrium.  Cosmic phase transitions provide an excellent
opportunity to create out-of-equilibrium effects, and phase
transitions are central to a wide variety of baryogenesis scenarios,
from the GUT scale all the way down to the electroweak
scale.  Some scenarios involve topological defects, while
others involve other out-of-equilibrium effects.  A more extensive
discussion of baryogenesis (including the connection with phase
transitions) can be found in section \ref{P48BG}.

{\bf Phase transitions and inflation}
As noted in subsection \ref{P48PTOver}, phase transitions were crucial
in creating the idea of cosmic inflation.  They provided the first
specific mechanism for how the Universe could enter an inflationary
state, and also led to the monopole problem, which stimulated a
fresh thinking about early Universe cosmology.  Phase
transitions continue to play a central role in the development of
the inflationary scenario, and bubbles produced in a higher
dimensional phase transition are at the core of a fascinating new
alternative to inflation\cite{Bucher:2001it}.

{\bf Cosmic Rays}
Cosmic rays are the most energetic particles observed, and they  have energies
almost a billion
times greater than particles in the Tevatron.
The origin of these particles remains a mystery. Defects formed in
cosmic phase transitions carry energy densities set by the energy
scale of the phase transition, which could be upwards of
$10^{16}GeV$. Topologically stable defects would persist, to some
degree at least, until the present day, and could perhaps produce ultra high
energy cosmic rays. Thus cosmic rays could be
providing us with a window on symmetry breaking at ultra-high
energies.  (Cosmic rays in general are discussed at length in the
report of group P4.5.)

{\bf Gravity Waves}
Perhaps the most ambitious frontier of physics is the pursuit of
gravity wave detection.  Because the energy scales for cosmic defects
can be extremely high, cosmic defects can be a significant source of
observable gravitational waves. In fact, for topologically stable
defects, the emission of gravity waves is often the only decay channel,
and significant amounts of gravity waves are produced. (See the
reports of the P4.3 and P4.6 groups for further discussion.)

{\bf Cosmic Magnetic Fields}
Magnetic fields of $10^{-6}$ Gauss are common within galaxies, and
extragalactic magnetic fields are also present, although at lower
strength.  The origin of
these fields, and especially of the small ``seeds'' that could be
amplified by astrophysical processes, remains a mystery.  One very
interesting possibility is that primordial magnetic fields were
generated in a cosmic phase transition\cite{Vachaspati:1991,Berera:1998hv,1998tsra.conf...88O}.

{\bf Exotic Objects}
Phase transitions produce dramatic out-of-equilibrium effects, and
cosmic phase transitions can generate a wide range of
exotic  objects. Such objects could be a cosmological disaster (such
as domain walls or monopoles), or they could help explain significant
phenomena (for example, ``WIMP-zillas'', a candidate for the Dark Matter, could
have formed in a cosmic phase transition\cite{Kolb:1998ki}).

\section{Baryogenesis}
\label{P48BG}

All observations show that the universe is
baryon-antibaryon asymmetric, and that there is negligible primordial
antimatter in our observable universe \cite{Steigman:1976ev}. To obtain
a quantitative measure of this asymmetry we look to the standard
cosmological model. One of the major successes of cosmology is an
accurate prediction of the abundances of all the light elements; a
calculation which requires a single input parameter, the {\it
baryon to entropy ratio}
\begin{equation}
\eta \equiv \frac{n_B}{s} = \frac{n_b-n_{\bar b}}{s} \ ,
\end{equation}
where $n_b$ is the number density of baryons, $n_{\bar b}$ is that
of antibaryons, and $s$ denotes the entropy density. If one
compares calculations of elemental abundances with observations,
then there is agreement between these numbers if
\begin{equation}
1.5\times 10^{-10} < \eta < 7\times 10^{-10} \ . \label{nucleo}
\end{equation}
This number is an input parameter in the standard model of
cosmology, and it is one of the goals of particle cosmologists to
understand its origin from particle physics.

In 1968, Sakharov \cite{Sakharov:1967dj} identified the conditions necessary for a
particle physics theory to generate any asymmetry between baryons
and antibaryons. These are violations of the baryon number (B),
the charge (C) and charge-parity (CP) symmetries, and a departure
from thermal equilibrium. One can imagine (any many have) a great
number of ways in which this combination of circumstances could be
arranged in the context of particle physics in an expanding
universe. However, it seems fair to devote most attention to those
mechanisms which arise as a natural consequence of particle
physics theories proposed for other compelling phenomenological
reasons. In addition, those mechanisms amenable to experimental
tests in the near future, particularly electroweak baryogenesis,
deserve our immediate attention (for reviews
see \cite{Riotto:1999yt,Riotto:1998bt,Trodden:1998ym,Rubakov:1996vz,Cohen:1993nk,Dolgov:1992fr}).

Certainly, the original suggestion that grand unified theories
(GUTs) may be responsible for the BAU is firmly in the first
category, but probably not in the second. Grand unification is an
attractive idea for understanding the origin of the standard
model, the apparent meeting of the running $SU(3)$, $SU(2)$ and
$U(1)$ couplings, and the quantization of charge. Further, baryon
number is naturally violated in GUT models because quarks and
leptons lie in the same representation of the grand unified gauge
group, and C and CP may also naturally be violated. The required
departure from thermal equilibrium must have an entirely
cosmological origin, and in this case it occurs because the
expansion rate of the universe at the GUT epoch is significantly
faster than the rate of particle interactions. However, despite
its attractive properties, there are an number of problems to be
overcome by such models. While this is not the place to provide a
detailed treatment of these, one is particularly relevant to our
subsequent discussion.

The electroweak theory itself violates baryon number (and to an
identical amount lepton number (L)) through an anomaly \cite{'tHooft:1976fv}. While this
is irrelevant at zero temperature (since the relevant phenomenon
is mediated by an instanton of large action, and hence has a close
to vanishing rate), at temperatures around or above the
electroweak scale such events are unsuppressed and copious \cite{Kuzmin:1985mm}. One
consequence of this is that if a GUT model does not produce a net
baryon minus lepton number (B-L) asymmetry rather than just a
baryon asymmetry, then anomalous electroweak interactions between
the GUT and electroweak scales will erase the asymmetry. However,
the presence of baryon number violation in the electroweak theory
at finite temperature suggests that this theory itself may be
capable of generating the
BAU \cite{Shaposhnikov:1986jp,Shaposhnikov:1987tw,Shaposhnikov:1988pf,Cohen:1987vi,Cohen:1988kt,Cohen:1990py,Cohen:1991it,Cohen:1991iu,Nelson:1992ab,Turok:1990in,Turok:1991zg}.
Of course, there are two other
Sakharov conditions to be satisfied. In the standard model the
condition of C violation is maximal and CP violation is present at
a small level (as evidenced in the Kaon system). However, even if
the level of CP violation were enough (which it is not) there is
insufficient departure from thermal equilibrium at the electroweak
scale, since the minimal electroweak phase transition is
continuous for Higgs masses is the experimentally allowed range.
This observation has led to the hope that the minimal
supersymmetric standard model (MSSM) may allow for electroweak
baryogenesis.

The behavior of the electroweak phase transition in the minimal
supersymmetric standard model is dependent on the mass of the
lightest Higgs particle, and the mass of the top squark. A variety of
analytical \cite{Carena:1996wj,Delepine:1996vn,Espinosa:1996qw,Bodeker:1997pc,Losada:1997ju,Farrar:1997cp,deCarlos:1997ru,Carena:1998ki,Losada:1998at}
and
lattice \cite{Laine:1996ms,Laine:1998vn,Cline:1996cr,Laine:1998qk}
computations have  revealed that the
phase transition can be sufficiently strongly first order  in the
presence of a top squark lighter than the top quark. In order to
naturally suppress contributions to the $\rho$-parameter, and
hence preserve a good agreement with precision electroweak
measurements at LEP, the top squark should be mainly right handed.
This can be achieved if the left handed stop soft supersymmetry
breaking mass $m_Q$ is much larger than $M_Z$.

The preservation of the baryon number asymmetry requires  the
order parameter $\langle \phi(T_c)\rangle /T_c$ to be larger than
one. In order to obtain values of $\langle \phi(T_c)\rangle/T_c$ larger
than one, the Higgs mass must take small values, close to the
present experimental bound. Hence,  small values of $\tan\beta$
are preferred. The larger the left handed stop mass, the closer to
unity $\tan\beta$ must be. This implies that the left handed stop
effects are likely to decouple at the critical temperature, and
hence that $m_Q$ mainly affects the baryon asymmetry through the
resulting Higgs mass.  A detailed analysis,  including all
dominant two-loop finite temperature corrections to the Higgs
effective potential and the non-trivial effects arising from
mixing in the stop sector, has been performed \cite{Carena:1998ki}, and
the region of parameter space for which MSSM electroweak baryogenesis
can happen identified. Taking into
account  the experimental bounds as well as the requirement of
avoiding dangerous color breaking minima, it was found that the
lightest Higgs should be lighter than about $105$ GeV, while the
stop mass may be close to the present experimental bound and must
be smaller than, or of order of, the top quark
mass \cite{Carena:1998ki,Laine:1998vn}. This lower
bound  has been essentially confirmed  by lattice simulations
\cite{Laine:1998qk}, providing a motivation for the search for Higgs and
stop particles at the Tevatron and future colliders.

The popularity of this idea is tightly bound to its testability.
The physics involved is all testable in principle at realistic
colliders. Furthermore, the small extensions of the model involved
to make baryogenesis successful can be found in supersymmetry,
which is an independently attractive idea, although electroweak
baryogenesis does not depend on supersymmetry.
The most direct
experimental way of testing this scenario is through the search
for the lightest Higgs.  In this sense, we are close to
knowing whether electroweak processes were responsible for the
BAU.

If the Higgs is found, the second test will come from the search
for the lightest stop at the Tevatron collider. If both
particles are found, the last crucial test will come from $B$
physics, more specifically, in relation to the CP-violating
effects.

Moreover, the selected parameter space leads to values of the
branching ratio ${\rm BR}(b\rightarrow s\gamma)$ different from
the Standard Model case. Although the exact value of this
branching ratio depends strongly on the value of the $\mu$ and
$A_t$ parameters, the typical difference with respect to the
Standard Model prediction is of the order of the present
experimental sensitivity and hence in principle testable in the
near future. Indeed, for the typical spectrum considered here, due
to the light charged Higgs, the branching ratio ${\rm BR}(b
\rightarrow s \gamma)$ is somewhat higher than in the SM case,
unless negative values of $A_t\mu$  are present. The crucial nature of
knowledge concerning CP violation in the $B$-sector for baryogenesis means that
the results of the BaBar \cite{babar},
BTeV \cite{btev}
Belle \cite{belle} and
LHCb \cite{lhcb} experiments, for example
the BaBar measurement of $\sin(2\beta)$ \cite{Aubert:2001nu} announced during the
Snowmass meeting, will be particularly useful.

We now turn to a third baryogenesis scenario, that has received a
lot of attention. This mechanism was introduced by Affleck and
Dine (AD) \cite{Affleck:1985fy} and involves the cosmological
evolution of scalar fields carrying baryonic charge. These scenarios are most
naturally implemented in the context of
supersymmetric models (e.g. \cite{Dine:1996kz}).
Consider a colorless, electrically neutral combination of quark
and lepton fields. In a supersymmetric theory this object has a
scalar superpartner, $\chi$, composed of the corresponding squark
${\tilde q}$ and slepton ${\tilde l}$ fields.

Now, an important feature of
supersymmetric field theories is the existence of ``flat
directions" in field space, on which the scalar potential
vanishes. Consider the case where some component of the field
$\chi$ lies along a flat direction. By this we mean that there
exist directions in the superpotential along which the relevant
components of $\chi$ can be considered as a free massless field.
At the level of renormalizable terms, flat directions are generic,
but supersymmetry breaking and nonrenormalizable operators lift
the flat directions and sets the scale for  their potential.

During inflation it is natural for the $\chi$ field to be displaced
from the position $\langle\chi\rangle=0$, establishing the initial
conditions for the subsequent evolution of the field. An important
role is played at this stage by baryon number violating operators
in the potential $V(\chi)$, which determine the initial phase of
the field. When the Hubble rate becomes of the order of the
curvature of the potential, the condensate starts
oscillating around its minimum. At this time,
$B$-violating terms in the potential are of comparable importance
to the mass term, thereby imparting a substantial baryon number to
the condensate. After this time, the baryon number violating
operators are negligible so that, when the baryonic charge of
$\chi$ is transferred to fermions through decays, the  net baryon
number of the universe is  preserved by the subsequent
cosmological evolution.

The challenges faced by Affleck-Dine models are combinations of
those faced by the GUT and electroweak ideas. In particular, it is
typically necessary that $B-L$ be violated along the relevant directions
and that there
exist new physics at scales above the electroweak. If
supersymmetry is not found, then it is hard to imagine how the
appropriate flat directions can exist in the low energy models.

Of all models for baryogenesis, the electroweak scenario has received
most attention.
Electroweak baryogenesis is such an attractive idea because it is
testable and uses physics that is already there for a good
particle physics reason. If the model is successful, it is a triumph of
the particle physics/cosmology union. If not, our
primary attention should be focused on models with the same
properties. It is possible that Affleck-Dine models may fit the bill, or
that the discovery of neutrino masses is telling us something useful
about the direction to go

\section{Fundamental Physics and Cosmology}
\label{P48FP}

Since the last Snowmass meeting, tremendous progress has been made
in understanding the non-perturbative structure of string theory
and the theory into which it has evolved - M-theory. These recent
advances may be at last opening the door for a serious approach to
analyzing the earliest times in the universe. Since the early
universe is a hot, dense, highly energetic place and time, we
expect a non-perturbative understanding of quantum effects in
gravity to be essential to an analysis of cosmology in such an
environment. For those who believe our basic framework for
addressing these questions is in place, the recent explosion of
interest in extra-dimensional physics should act as a cautionary
tale. Solid tests of our cosmological model go back only as far as
the epoch of primordial nucleosynthesis, at which the temperature
of the universe was still only a few MeV. If we are to gain a
quantitative understanding of earlier epochs, it will be necessary
to develop a consistent approach to theoretical analyses of the
early universe, and innovative new ideas for probing the nature of
space and time at these early times. While most ideas in this field
are wildly speculative at the present, we nevertheless present some
interesting first steps here.

\subsection{Finite Temperature String / M-Theory and Cosmology}
If we accept string / M-Theory as the correct theory of everything,
then it must hold the key to the most fundamental problems in cosmology.
Perhaps the most obvious modification to the
standard equations of cosmology is that gravity is no longer described purely
by general relativity. In particular, Einstein's theory is modified by the
appearance of a dilaton field, related to the compactification of the theory,
and by higher derivative terms in the action, at nonzero order in the string
coupling constant. There have been a number of attempts to use these
modifications to address the origin of the hot big bang phase of the universe,
the issue of the initial singularity, and even the origin of the number of
macroscopic dimensions we observe.

In the Brandenberger-Vafa scenario \cite{Brandenberger:1989aj,Tseytlin:1992xk}, it is assumed
that the fundamental physics respects a T-duality, interchanging large and small radii
of a toroidal compactification. This has two particularly interesting implications for
cosmology. First, since the small radius of the universe limit is equivalent to the large
radius limit, there is no big-bang singularity in the usual general relativistic sense.
Second, the dynamics of string winding modes constrains the number of dimensions that
may become macroscopic. Imposing T-duality on the modified Einstein equations results in
string winding around a particular direction preventing the expansion of that dimension.
In the early universe, at high temperatures, it is argued that one should think of a gas
of strings (and branes in an extended picture \cite{Alexander:2000xv,Brandenberger:2001kj})
and their modes. In more than three spatial
dimensions, string winding modes cannot annihilate, and hence only as many as three dimensions
may decompactify in this picture. Although there are a number of problems with this picture
(such as the need for non-trivial one-cycles to wrap around, which do not exist in typical
Calabi-Yau compactifications) this provides an interesting possiblity for explaining features
of the universe that are inaccessible to our lower energy effective theories. In general, a
careful analysis of the implications of finite temperature effects in string theory seems an
interesting research avenue to be pursued.

\subsection{Extra Dimensions and Cosmology}
The flurry of interest in extra dimensional physics
\cite{Antoniadis:1990ew,Lykken:1996fj,Arkani-Hamed:1998rs,Antoniadis:1998ig,Randall:1999ee}
has led to a
number of interesting proposals for modifying cosmology at early
times. While one must be careful not to interfere with the
successful predictions of the standard cosmology
\cite{Arkani-Hamed:1998nn,Hall:1999mk,Cullen:1999hc}, the evolution of
the universe at the earliest times must be very different in these
models from that expected in the usual $3+1$ dimensional
framework \cite{Arkani-Hamed:1999gq}. When confronted with a new model for the early
universe, cosmologists typically ask themselves the following
three fundamental questions: 1) Is there a new way to address the
horizon and flatness problems in this picture? 2) Is there a way
to understand the observed size of the cosmological constant in
this picture? 3) Is there a new mechanism for generating the
density and temperature fluctuations seen in large scale structure
and the CMB? As far as the first point goes, there have been
several suggestions. In ref~\cite{Starkman:2000dy,Starkman:2001xu} it was suggested that if
the universe is the direct product of a (3+1)-dimensional FRW
space and a compact hyperbolic manifold \cite{Kaloper:2000jb}, the decay of massive
Kaluza-Klein modes leads to the injection of any initial bulk
entropy into the observable (FRW) universe. This can act to smooth
out any initial inhomogeneities in the distribution of matter and
of 3-curvature sufficiently to account for the current homogeneity
and flatness of the universe. In ref~\cite{Khoury:2001wf} it has been
suggested that the true vacuum of the universe is a BPS state in
heterotic M-theory. The fields describing our universe live on a
brane in this space and the hot expanding phase that we know as
the big bang arises due to an instanton effect \cite{Khoury:2001bz} in which a new
brane nucleates, travels across one of the extra dimensions and
collides with our brane, depositing its energy there. The eternal
nature of the cosmology and the special, flat nature of the BPS
state lead to a flat and homogeneous FRW cosmology on our brane
after the collision. While there may be existing fine-tunings in
the above models, their feasibility and implications for cosmology
are under investigation.

The cosmological constant problem remains unaddressed in the above
scenarios. However, a particular scheme, the so-called self-tuning
model \cite{Arkani-Hamed:2000eg,Kachru:2000hf},
has been proposed in the bane world context to address this
issue. The cosmological constant problem arises in a simple sense
because space-time is sensitive to the presence of vacuum energy.
In the self-tuning picture, the effective theory of gravity on our
brane is such that it does not respond to the presence of vacuum
energy. This occurs because of a careful choice of coupling
between the matter fields living on the brane and in the
5-dimensional bulk. As a result, whatever the value of the
cosmological constant, or however it may change, the
$3+1$-dimensional metric remains unaffected by it. However, there
are a number of unresolved questions regarding the self-tuning
models. First, there seems to be a singularity in these models,
the possible resolution of which is yet to be understood
\cite{Forste:2000ps,Csaki:2000wz,Horowitz:2000ds,Grinstein:2000fn,Zhu:2000bt,Barger:2000wj,Binetruy:2000wn,Maeda:2000wr,Mendes:2000wu,Kakushadze:2000ix,Kennedy:2000vq,Kim:2000mc,Kim:2001ez,Brax:2001fh}.
Second,
it seems quite difficult to reconcile the self-tuning picture with
what is known about the cosmology of our universe. In particular,
it seems necessary to modify the Friedmann equation on our
brane \cite{Binetruy:1999ut,Csaki:1999jh,Cline:1999ts,Binetruy:1999hy,Shiromizu:1999wj,Flanagan:1999cu}
, which can lead to problems with big bang nucleosynthesis \cite{Carroll:2001zy}.

Finally, the issue of density and temperature perturbations is a
highly quantitative challenge to any new theory of the early
universe. The precision measurements of the temperature
fluctuations in the CMB, and in particular the observation of the
second (and perhaps third) acoustic peaks in the power spectrum
now provide solid evidence for a scale-free adiabatic spectrum of
initial fluctuations, consistent with that predicted by inflation.
It is possible that inflation in the brane world (e.g. see
\cite{Dvali:1998pa,Burgess:2001vr,Burgess:2001fx}) could be
responsible for this, but the question of whether any other
mechanism could be responsible for these observations is a
particularly pressing one for particle cosmology. Recently it was
claimed that one of the scenarios mentioned above, the ekpyrotic
scenario \cite{Khoury:2001wf}, may be able to generate the
necessary perturbations. However, at present this is a hotly
debated topic
\cite{Lyth:2001pf,Brandenberger:2001bs,Hwang:2001ga,Lyth:2001nv,Khoury:2001zk},
and the ultimate outcome is unclear. Certainly, if this claim is
true, the ekpyrotic scenario will have earned further careful
study.

\subsection{de-Sitter Space as a Solution to M-Theory}
One interesting development to come out of the particle
physics-cosmology interface is the role of de-Sitter space in
M-theory. If inflationary theory is correct, then the universe
must go through an accelerating (perhaps quasi-de-Sitter) phase at
early times. However, the successes of the standard cosmology (not
to mention our presence in the universe today) imply that this was
a transient phase. However, the recent observations of type IA
supernovae \cite{Perlmutter:1998np,Riess:1998cb} point to a second
accelerating epoch, beginning at the present time. Taking this
data at face value, there are two interesting possibilities: the
acceleration could be caused by a small non-zero cosmological
constant, or by some type of energy that redshifts sufficiently
slowly as to cause acceleration, but that will eventually cease to
act. Let us focus on the former possibility. If there exists a
true cosmological constant in the universe, then the late-time
space-time will approach de-Sitter space. This would seem to imply
that de-Sitter space was a vacuum of the underlying theory. In the
context of string theory this may be a problematic conclusion.
Several authors \cite{Hellerman:2001yi,Fischler:2001yj} have
recently pointed out that de-Sitter space seems to be incompatible
with string theory, at least at the level of perturbation theory.
Other arguments, based on upper bound on entropy in de-Sitter
space\cite{Banks:2000fe} also challenge the viablity of string
theory (or any theory with infinite degrees fo freedom) in a
universe with a real cosmological constant. If, as measurements of
the equation of state of the dark energy are refined, and string
(or M) theory matures and its non-perturbative structure is
understood, this tension remains, this may be a a way for
cosmology to constrain our fundamental theories.

\subsection{Transplanckian Physics and Cosmology}
\label{P48TransPlanck} Quantum effects during inflation may be the
origin of the temperature fluctuations observed today in the CMB.
If so, then scales observed today in large scale structure
originated at smaller than Planck scales at the beginning of
inflation. Thus, cosmological observations may reveal the
structure of physics at sub-Planckian distances. The problem
arises because, since there are typically many more than 60
e-foldings in inflationary models, we must extrapolate the
weakly-coupled field theories we understand into regimes in which
the approximation is no longer valid in order to extract
information about the fluctuations. Although we should not trust
this procedure, there is at present no way to calculate the
expected dispersion relation of fluctuation modes from first
principles in any fundamental theory. To make progress, several
authors
\cite{Martin:2000xs,Brandenberger:2000wr,Niemeyer:2000eh,Niemeyer:2001qe,Kempf:2000ac,Kempf:2001fa,Easther:2001fi,Hui:2001ce,Easther:2001fz}
have adopted a phenomenological approach, trying different
modifications to the dispersion relation at small scales and
studying the effects on the expected spectrum of fluctuations.

In general, it proves quite difficult to make short-scale
modifications that lead to large effects in cosmological
observables (although it has been suggested that these modes might
provide a new origin for dark energy \cite{Bastero-Gil:2001nu}).
However, this may prove to be a useful technique for testing
fundamental physics. In particular, the effects of transPlanckian
physics are expected to alter the ratio of the spectrum of scalar
to tensor perturbations generated during inflation. In a large
class of inflationary models the evolution is dominated by a
single field. In that case this ratio is a fixed known number, and
deviations from it must signal some departure from the standard
picture. Whether the specific signals predicted from short-scale
modifications of the dispersion relation can be isolated from
other possible effects remains to be seen. Nevertheless, we
consider any possible cosmological tests of our most fundamental
theories an interesting avenue to pursue.

%
%

%
%

\begin{acknowledgments}
MT thanks Toni Riotto for useful discussions on baryogenesis. AA is
supported by DOE grant DE-FG03-91ER40674.  The work of MT is supported
by the NSF under grant PHY-0094122
\end{acknowledgments}

\bibliography{P4_8_EarlyUniverse}

\end{document}